\documentclass[aps,twocolumn,showpacs,preprintnumbers,prb,superscriptaddress]{revtex4-2}
\usepackage{graphicx}
\usepackage{dcolumn}
\usepackage{tikz}
\usepackage{bm}
\usepackage{amssymb}
\usepackage{framed}
\usepackage{amsmath}
\usepackage{appendix}
\usepackage{hhline}
\usepackage[dvipsnames]{xcolor}
\definecolor{bluegreen}{rgb}{0,0.2,0.8}
\usepackage{subfigure,amsmath,verbatim,moreverb}
\usepackage{subcaption}
\usepackage{tabularx}
\usepackage{adjustbox}
\usepackage{booktabs}
\usepackage[dvipsnames,table]{xcolor}
\usepackage{latexsym}
\usepackage{epsf}
\usepackage{float}
\usepackage[breaklinks=true,colorlinks,citecolor=blue,linkcolor=blue,urlcolor=blue]{hyperref}
\usepackage[normalem]{ulem}
\usepackage{multirow}
\usepackage{siunitx}
\usepackage{threeparttable}
\usepackage[T1]{fontenc}
\usepackage{array}

\usepackage{etoolbox}
\AtBeginEnvironment{align}{\setcounter{subeqn}{0}}
\newcounter{subeqn}

\newcommand{\kcal}{kcal\,mol$^{-1}$}
\newcommand{\rtscan}{r$^2$SCAN}
\newcommand{\mlrtscan}{ML-r$^2$SCAN}

\begin{document}

\title{What Does a Semilocal Machine-Learning Correction Actually Learn?\\
Size-Dependent Errors across Four Parent Functionals}

\author{Abhishek Bhattacharjee}
\email{abhishek.bhattacharjee@niser.ac.in}
\affiliation{School of Physical Sciences, National Institute of Science
Education and Research, An OCC of Homi Bhabha National Institute,
Jatni 752050, India}
\author{Kishan Kumar Mohanta}
% \email{kishankumar.mohanta@niser.ac.in}
\affiliation{School of Physical Sciences, National Institute of Science Education and Research, An OCC of Homi Bhabha National Institute, Jatni 752050, India}
\author{Subrata Jana}
%\email{subrata.jana@umk.pl, subrata.niser@gmail.com}
\affiliation{Institute of Physics, Faculty of Physics, Astronomy and Informatics, Nicolaus Copernicus University in Toru\'n, ul. Grudzi\k{a}dzka 5, 87-100 Toru\'n, Poland}
\author{Prasanjit Samal}
%\email{psamal@niser.ac.in}
\affiliation{School of Physical Sciences, National Institute of Science Education and Research, an OCC of Homi Bhabha National Institute, Bhubaneswar 752050, India}

\begin{abstract}

Machine-learning (ML) corrections to density functional approximations (DFAs) offer a route to improving electronic-structure predictions while retaining the efficiency of semilocal functionals. An important question is whether such corrections learn transferable improvements or instead compensate for errors specific to the parent functional. Here, we address this question by applying the semilocal ML correction of Wang \textit{et al.} [J.~Chem.~Phys.~\textbf{158}, 154107 (2023)] to PBE, B3LYP, SCAN, and r$^2$SCAN, while keeping the network architecture, loss function, training set, and optimization protocol unchanged. We find that the learned correction develops a systematic size-dependent contribution to atomization energies for all four parents, with  $|r|\geq0.9999$ along the $n$-alkane series, where r is the Pearson coefficient of the linear fit. Its magnitude and sign depend systematically on the parent: the correction partially compensates the size-dependent errors of PBE and B3LYP, but introduces substantial size-dependent contributions for SCAN and r$^2$SCAN, whose parent errors show little size dependence. In contrast, ionization potentials, electron affinities, and isomerization energies are largely unaffected. Spatial decomposition further reveals distinct origins of this extensive contribution, with bonding regions dominating for B3LYP, SCAN, and r$^2$SCAN and core regions dominating for PBE. These results show that the global-energy ML correction is strongly dependent on the error structure of its parent DFA, highlighting size-dependent error accumulation as a critical consideration in the assessment and development of ML-enhanced density functionals.

\end{abstract}

\maketitle

% =====================================================================
\section{Introduction}
% =====================================================================

%{\textcolor{red}{specify appendix number clearly in main text. For example in main text define "This is shown in Appendix S1~\ref{appendix}." }}

The accuracy of Kohn--Sham density functional theory (KS-DFT)
~\cite{HK} depends critically on the quality of the exchange--correlation
(XC) functional~\cite{KS,PerdewKurth2003}, which must be approximated in practical
calculations. Five decades of functional development have produced a
hierarchy of increasingly sophisticated approximations based on physical
constraints and known limits~\cite{becke2014perspective}. These density
functional approximations (DFAs) form the basis of much of modern
electronic-structure theory. Machine learning (ML) has recently emerged as
a route to reducing systematic errors of existing DFAs while retaining the
KS-DFT framework and its computational advantages. ML-based approaches
range from fully machine-learned neural-network functionals
~\cite{nagai2020completing,dm21,skala,polak2025real} to ML corrections constructed on top
of established parent DFAs~\cite{wang2023,wang2022,deepks,zheng2025machine}.
The latter approach is particularly attractive because the parent DFA
retains its known physical constraints and computational structure, while
the ML model is required to describe only the residual error.

An ML correction, however, need not represent a transferable physical
contribution. An additive semilocal correction can improve the accuracy of a
particular parent DFA either by learning a more general component of the
XC functional or by compensating for errors specific to that parent.
These two possibilities cannot be distinguished from the accuracy of a
single ML-corrected functional. A direct way to examine this question is to
keep the ML architecture, training procedure, and training data fixed while
changing the parent DFA. If the learned correction represents a transferable
component of the XC functional, similar improvements should be expected
across different parents. If instead it primarily compensates for
parent-specific errors, its effect should depend strongly on the parent
functional.

Wang \textit{et al.}~\cite{wang2023} introduced an ML correction to B3LYP~\cite{becke,becke88,lyp,vwn}
based on a semilocal mapping from density descriptors to an additive XC
energy correction. The model was trained self-consistently using global
energetic information from three small molecules and was shown to reduce
the mean absolute deviation of atomization energies for the G2 test set
~\cite{curtiss1997assessment,curtiss1998assessment}. This construction provides a useful framework for studying
whether a fixed ML correction behaves consistently when applied to
different parent DFAs.

Here, we systematically examine this question by applying the ML-DFA
correction scheme of Ref.~\cite{wang2023} to four parent functionals
spanning three rungs of Jacob's ladder\cite{perdew2001jacob,perdew2005prescription}: the GGA PBE~\cite{pbe}, the hybrid
GGA B3LYP~\cite{becke,becke88,lyp,vwn}, and the meta-GGAs SCAN~\cite{scan}
and r$^2$SCAN~\cite{r2scan}. We evaluate atomization energies using three
test sets and further examine ionization potentials, electron affinities,
and isomerization energies. This allows us to separate improvements that
are specific to a particular energetic property from more systematic
changes in the accuracy of the parent DFA.

Our results show that the learned correction develops a systematic
size-dependent contribution for all four parent functionals. For PBE and
B3LYP, this contribution leads to improvements consistent with the behavior
reported previously~\cite{wang2023}. For SCAN and r$^2$SCAN, however, the
same ML construction introduces a size-dependent error that degrades the
performance of the parent functionals. Thus, the effect of the ML
correction is strongly dependent on the parent DFA. These results show
that an ML correction that improves one parent functional cannot, in
general, be assumed to represent a transferable correction applicable to
other DFAs.

% That is what we do here. We apply the scheme of Ref.~\cite{wang2023}
% unchanged to four parents spanning three rungs of the Jacob's ladder
% hierarchy --- the GGA PBE~\cite{pbe}, the hybrid GGA
% B3LYP~\cite{becke,becke88,lyp,vwn}, and the meta-GGAs SCAN~\cite{scan} and
% \rtscan~\cite{r2scan} --- and benchmark each on atomisation energies,
% ionisation potentials and electron affinities. The answer is unambiguous,
% and it is not the favourable one: the correction learns a size-extensive
% term whose magnitude and usefulness are set by the parent rather than by the
% network.

\section{Relation to prior ML work}
\label{sec}
Size-dependent errors are a long-standing issue in DFAs. Errors in semilocal and hybrid functionals can
accumulate with system size, leading to increasing errors in extensive
quantities such as total energies and thermochemical properties
~\cite{SizeErr-B3lyp_2000,SizeErr-B3lyp_2006}. For ML corrections
to DFAs, Wang \textit{et al.}~\cite{wang2022} showed that systematic
pointwise errors in a local ML correction can accumulate as the system size
increases. Using the G3-3 heats-of-formation subset~\cite{zhao2006new} of the G3/99
database~\cite{SizeErr-B3lyp_2000}, commonly denoted as G3-HOF, they found that
the PBE error increases approximately linearly with the number of
non-hydrogen atoms, with a slope of 9.3~kcal\,mol$^{-1}$ per non-hydrogen atom. Their ML-PBE
correction reduced this slope to approximately 2.9~kcal\,mol$^{-1}$ per non-hydrogen atom,, corresponding to a
reduction of about $69\%$~\cite{wang2022}. Subsequent work introduced a
global-loss ML-DFA scheme, denoted here as the ``$-g$'' construction, based
on molecular energies and energy differences~\cite{wang2023}.

Our central comparison is different. We examine whether an ML correction
preserves the size behavior of its parent DFA. For SCAN and r$^2$SCAN, the
errors of the parent functionals show little systematic growth with system
size, whereas their ML corrections develop a systematic size-dependent
error (Fig.~\ref{fig:accum}). Thus, the observed
size dependence is not simply inherited from the parent DFA; it is
introduced, at least in part, by the ML correction. This distinction is
important because good size behavior of the parent functional does not
guarantee the same behavior after ML correction.

This observation should not be interpreted as a general limitation of ML
corrections to meta-GGAs. Nagai \textit{et al.}~\cite{nagai2022pcnn}
introduced a neural-network (NN) correction to SCAN using H$_2$O, NH$_3$, and
NO, together with density-distribution information, and reported improved
atomization energies for a test set of 144 molecules. Their construction
differs from the present ML-DFA framework in its training targets,
descriptors, and imposed constraints. The size-dependent behavior observed
here should therefore be associated with the specific ML-DFA construction
and training protocol considered in this work, rather than with ML
corrections to meta-GGAs in general.

The ``$-g$'' and ``$-p$'' constructions considered in the literature also
differ in how the ML correction is trained. The ``$-g$'' scheme uses a global loss based on molecular energetic quantities~\cite{wang2023}, whereas the ``$-p$'' construction uses pointwise training of the absolute
XC-energy correction, together with a larger training set and
species-dependent weighting~\cite{an2025mitigating}. The latter also uses the
meta-GGA-related descriptors $z$, $\alpha$, and $t^{-1}$.
Because several aspects of the two constructions change simultaneously,
their individual effects cannot be separated without dedicated ablation
studies. In particular, the present comparison does not establish whether
the observed size dependence arises from the loss function, training-set
composition, descriptors, network architecture, or their combination.

The role of the training target is nevertheless relevant. The ML-PBE
approach of Wang \textit{et al.}~\cite{wang2022} used pointwise information together with
atomic total energies and explicitly monitored error accumulation with
system size. More recent work on ML-corrected B3LYP has
also emphasized training on absolute XC energies and the limitations of
relying on error cancellation between different species~\cite{an2025mitigating}.
These results motivate treating the training objective as one possible
factor controlling the size transferability of an ML correction.

More generally, size transferability has been considered explicitly in
other ML-based density-functional constructions. For example, DeePKS~\cite{deepks}
incorporates locality into its model construction to facilitate application
to systems larger than those used for training. Our results
suggest that size extensivity should likewise be considered when designing
and training ML corrections, rather than being treated only as a property
to be checked after training.

\section{Theory and Methods}
% =====================================================================

\subsection{ML-corrected Functional form}

The ML-corrected XC energy is defined as in Ref.~\cite{wang2023},
\begin{equation}
E_{\mathrm{XC}}^{\mathrm{ML-DFA}}
=
E_{\mathrm{XC}}^{\mathrm{DFA}}
+
\int d\mathbf{r}\,
\rho(\mathbf{r})\,
\Delta\epsilon_{\mathrm{XC}}^{\mathrm{ML}}(\mathbf{r}),
\label{eq:mlxc}
\end{equation}
where $\Delta\epsilon_{\mathrm{XC}}^{\mathrm{ML}}$ is represented by a fully
connected NN with architecture
$3\times20\times20\times20\times1$ and sigmoid activation functions. The
network takes three semilocal descriptors as inputs,
\begin{equation}
r_s =
\left(\frac{4\pi}{3}\right)^{-1/3}\rho^{-1/3},
\qquad
\zeta =
\frac{\rho_\uparrow-\rho_\downarrow}{\rho},
\label{eq:desc}
\end{equation}
and
\begin{equation}
s =
\frac{|\nabla\rho|}
{2(3\pi^2)^{1/3}\rho^{4/3}}.
\label{eq:desc2}
\end{equation}
The output of the final layer is scaled by $10^{-3}$, and the
940-dimensional parameter vector is constrained to the interval
$[-20,1]$, following Ref.~\cite{repo}.

Because the ML correction depends only on $\rho$ and $\nabla\rho$, it
contains no explicit dependence on the kinetic-energy density $\tau$.
Consequently, when the parent DFA is a meta-GGA, all explicit $\tau$
dependence of the total XC energy originates from the parent functional.
The ML contribution therefore modifies only the $\rho$- and
$\nabla\rho$-dependent terms in the XC potential. In the implementation,
the corresponding derivatives with respect to the density variables are
obtained by automatic differentiation, while the $\tau$ derivative is
retained entirely from the parent DFA. Introducing $\tau$, or the
iso-orbital indicator $\alpha$ constructed from $\tau$
~\cite{beckeedgecombe,beckeroussel}, as an additional ML descriptor would
define a different model and is not considered here. Such kinetic-energy-
density information has, however, been shown to improve ML functionals
constructed directly from scratch~\cite{nagai2020completing}.

\subsection{Loss function and training}

We use the loss function given by Eq.~(5) of Ref.~\cite{wang2023},
\begin{align}
L ={}&
\frac{1}{N_{\mathrm{AE}}}
\sum_i
\frac{
\left|
\mathrm{AE}^{\mathrm{ML}}_i
-
\mathrm{AE}^{\mathrm{ref}}_i
\right|
}{
\left|
\mathrm{AE}^{\mathrm{DFA}}_{\mathrm{H_2O}}
\right|
}
\nonumber\\
&+
\alpha
\frac{1}{N_{\mathrm{TE}}}
\sum_j
\frac{
\left|
\mathrm{TE}^{\mathrm{ML}}_j
-
\mathrm{TE}^{\mathrm{ref}}_j
\right|
}{
\left|
\mathrm{TE}^{\mathrm{DFA}}_{\mathrm{H_2O}}
\right|
},
\label{eq:loss}
\end{align}
where $N_{\mathrm{AE}}=3$, $N_{\mathrm{TE}}=7$, and $\alpha=0.16$. The
normalization factors are evaluated separately for each parent DFA using
the corresponding uncorrected DFA energies. Thus, the same value of
$\alpha$ is used for all four models, while the absolute normalization of
the two terms is determined independently for each parent functional.

The training set contains the atomization energies of H$_2$O, C$_2$H$_2$,
and SO$_2$, together with the total energies of these three molecules and
the isolated H, C, O, and S atoms. These ten training quantities are the
same as those used in Ref.~\cite{wang2023}. The total-energy term prevents
the ML correction from reducing atomization-energy errors primarily through
cancellation between molecular and atomic total-energy errors.

The network parameters are optimized using the particle-swarm optimization
implementation~\cite{pso} in \texttt{nevergrad} \cite{nevergrad} version 0.4.3, with the
parameters $\omega=0.9$, $\phi_p=0.95$, and $\phi_g=0.9$, a population of
100, and a total budget of 3000 function evaluations. The swarm is
initialized at the origin, for which
$\Delta\epsilon_{\mathrm{XC}}^{\mathrm{ML}}=0$ and the ML-corrected
functional therefore reduces exactly to the parent DFA. Additional
implementation details are provided in the Appendix~\ref{App:implementaion}.

The optimization budget corresponds to 30 generations for a population of
100 in a 940-dimensional parameter space. Since each model is trained only
once, we do not assess convergence to the global minimum of
Eq.~\eqref{eq:loss} or the statistical variation of the resulting
parameters with different random seeds. We retain the particle-swarm
optimization protocol in order to reproduce the training procedure of
Ref.~\cite{wang2023}.

\subsection{Computational details}
\label{sec:comp}

All calculations are performed using PySCF~\cite{pyscf} with the
Def2-TZVPD basis set~\cite{basis}. We use integration-grid level 5 with
unpruned grids throughout. Pruning is disabled because SCAN-family
functionals can exhibit significant sensitivity to the numerical
integration grid~\cite{rscan}.

The supplementary material of Ref.~\cite{wang2023} does not report results
for SCAN or \rtscan{}, so the corresponding regression check available for
PBE and B3LYP cannot be performed directly for the meta-GGA parents. As a
validation of our implementation, we therefore recalculated the plain
B3LYP results using the modified code. We obtain a MAD of
$0.025~\mathrm{kcal\,mol^{-1}}$ relative to the published B3LYP results,
with a maximum deviation of $0.05~\mathrm{kcal\,mol^{-1}}$. The resulting
B3LYP MAD relative to CCSD(T) \cite{raghavachari1989fifth, bartlett2007coupled,haunschild2012new} is $6.09~\mathrm{kcal\,mol^{-1}}$. For the
meta-GGA calculations, the custom functional interface was extended to
propagate the kinetic-energy-density ($\tau$) contribution correctly
through the XC potential. Passthrough and one-electron tests validating
this implementation are described in the Appendix~\ref{App:implementaion}.

Second-order SCF optimization is not available for the ML-corrected
functionals because the ML correction does not provide an XC kernel. For
the parent DFA calculations, we use a convergence hierarchy consisting of
DIIS~\cite{pulay1980convergence,pulay1982improved}, followed by level shifting and a second-order SCF procedure when
needed. For ML-corrected calculations that do not converge from the
initial guess, we restart the SCF calculation from the converged density
of the corresponding parent DFA.

\begin{table*}
\centering
\caption{Mean absolute deviations (MADs) from CCSD(T) reference values
for the four parent functionals and their ML-corrected counterparts,
using the Def2-TZVPD basis set. All values are in
$\mathrm{kcal\,mol^{-1}}$. The first three columns contain absolute
atomization energies (AE), while the remaining four contain relative
energies for which the molecular composition is unchanged between the two
states. G2-AE denotes the atomization energies of the G2 test set~\cite{curtiss1997assessment,curtiss1998assessment};
Alk-AE denotes the 19 atomization energies of alkanes (Alk)~\cite{mardirossian2017thirty}; and P6-AE denotes
the six atomization energies of the Platonic hydrocarbon cages (P6)~\cite{mardirossian2017thirty}.
G2-IP and G2-EA denote ionization potentials and electron affinities,
respectively. ISO20~\cite{mardirossian2017thirty} contains 20 isomerization energies, while ISO-C~\cite{mardirossian2017thirty}
contains eight isomerization energies of C$_{20}$ and C$_{24}$.
The numbers in parentheses give the number of test cases. For G2-AE,
the plain functionals are evaluated for 148 species, whereas the
three training species are excluded for the ML-corrected functionals,
giving 145 test species.}
\label{tab:main}
\begin{ruledtabular}
\begin{tabular}{lccccccc}
& \multicolumn{3}{c}{Atomization energies}
& \multicolumn{4}{c}{Relative energies} \\
\cline{2-4}\cline{5-8}
Functional
& G2-AE & Alk-AE & P6-AE
& G2-IP & G2-EA & ISO20 & ISO-C \\
& (148/145) & (19) & (6)
& (46) & (33) & (20) & (8) \\
\hline
PBE        & 14.91 & 12.27 & 60.36 & 5.51 & 2.39 & 3.15 & 13.42 \\
ML-PBE     & 12.81 &  4.90 & 44.01 & 5.76 & 2.47 & 3.14 & 14.53 \\[2pt]
B3LYP      &  6.09 & 20.42 & 49.69 & 4.92 & 4.33 & 1.96 & 36.68 \\
ML-B3LYP   &  3.69 &  7.09 & 29.46 & 4.34 & 3.64 & 2.22 & 36.49 \\[2pt]
SCAN       &  3.11 &  1.42 &  4.81 & 6.47 & 5.88 & 2.44 &  9.76 \\
ML-SCAN    &  7.51 & 18.27 & 32.80 & 6.06 & 5.61 & 2.50 &  9.81 \\[2pt]
\rtscan    &  3.25 &  5.24 &  3.24 & 6.18 & 5.36 & 2.43 & 14.58 \\
\mlrtscan  &  5.36 &  7.24 & 18.99 & 5.88 & 5.05 & 2.39 & 14.06 \\
\hline
mean $|$change$|$
           & 2.75 & 9.89 & 20.08 & 0.39 & 0.34 & 0.09 & 0.47 \\
\end{tabular}
\end{ruledtabular}
\end{table*}

\begin{comment}
    
\begin{table*}%[t]
\centering
\caption{Mean absolute deviation from CCSD(T) references for the four parent
functionals with and without the ML correction, Def2-TZVPD, in \kcal.
The first three columns are absolute atomization energies; the last four are
relative energies, in which composition is unchanged between the two states
compared. G2-AE statistics for the ML-corrected functionals exclude the
three training species, so the plain rows cover 148 species and the
ML-corrected rows 145.}
\label{tab:main}
\begin{ruledtabular}
\begin{tabular}{lccccccc}
& \multicolumn{3}{c}{Atomization energies} & \multicolumn{4}{c}{Relative energies} \\
\cline{2-4}\cline{5-8}
Functional & G2 & Alk & P6-AE & G2-IP & G2-EA & ISO20 & ISO-C \\
           & (148/145) & (19)   & (6)   & (46)  & (33)  & (20)  & (8)   \\
\hline
PBE        & 14.91 & 12.27 & 60.36 & 5.51 & 2.39 & 3.15 & 13.42 \\
ML-PBE     & 12.81 &  4.90 & 44.01 & 5.76 & 2.47 & 3.14 & 14.53 \\[2pt]
B3LYP      &  6.09 & 20.42 & 49.69 & 4.92 & 4.33 & 1.96 & 36.68 \\
ML-B3LYP   &  3.69 &  7.09 & 29.46 & 4.34 & 3.64 & 2.22 & 36.49 \\[2pt]
SCAN       &  3.11 &  1.42 &  4.81 & 6.47 & 5.88 & 2.44 &  9.76 \\
ML-SCAN    &  7.51 & 18.27 & 32.80 & 6.06 & 5.61 & 2.50 &  9.81 \\[2pt]
\rtscan    &  3.25 &  5.24 &  3.24 & 6.18 & 5.36 & 2.43 & 14.58 \\
\mlrtscan  &  5.36 &  7.24 & 18.99 & 5.88 & 5.05 & 2.39 & 14.06 \\
\hline
mean $|$change$|$ & 2.75 & 9.89 & 20.08 & 0.39 & 0.34 & 0.09 & 0.47 \\
\end{tabular}
\end{ruledtabular}
\end{table*}
\end{comment}

% =====================================================================
\section{Results}
% =====================================================================

\subsection{Benchmark performance}
\label{sec:baselines}

Table~\ref{tab:main} summarizes the MADs relative to CCSD(T) \cite{raghavachari1989fifth, bartlett2007coupled,haunschild2012new} for the four
parent functionals and their ML-corrected counterparts. The effect of the
ML correction depends strongly on both the parent functional and the type
of energy difference considered.

For atomization energies, the ML correction improves PBE and B3LYP but
degrades SCAN and \rtscan{}. For G2-AE, the MAD decreases from
$14.91$ to $12.81~\mathrm{kcal\,mol^{-1}}$ for PBE and from
$6.09$ to $3.69~\mathrm{kcal\,mol^{-1}}$ for B3LYP. In contrast, the MAD
increases from $3.11$ to $7.51~\mathrm{kcal\,mol^{-1}}$ for SCAN and from
$3.25$ to $5.36~\mathrm{kcal\,mol^{-1}}$ for \rtscan{}. The difference
becomes more pronounced for the larger atomization-energy test sets. For
Alk-AE, the PBE and B3LYP MADs decrease from $12.27$ to
$4.90~\mathrm{kcal\,mol^{-1}}$ and from $20.42$ to
$7.09~\mathrm{kcal\,mol^{-1}}$, respectively, whereas the SCAN and
\rtscan{} MADs increase from $1.42$ to $18.27~\mathrm{kcal\,mol^{-1}}$
and from $5.24$ to $7.24~\mathrm{kcal\,mol^{-1}}$, respectively. The
contrast is even stronger for P6-AE: the SCAN MAD increases from $4.81$ to
$32.80~\mathrm{kcal\,mol^{-1}}$, while the \rtscan{} MAD increases from
$3.24$ to $18.99~\mathrm{kcal\,mol^{-1}}$.

The behavior of the parent functionals provides useful context for these
changes. SCAN and \rtscan{} have very small MSDs for the G2-AE set,
$-0.10$ and $+0.15~\mathrm{kcal\,mol^{-1}}$, respectively, compared with
$-5.63~\mathrm{kcal\,mol^{-1}}$ for B3LYP and
$+13.54~\mathrm{kcal\,mol^{-1}}$ for PBE. Thus, the small MSDs of the
meta-GGAs indicate that their atomization-energy errors are already
well balanced on average. Their non-negligible MADs of $3.11$ and
$3.25~\mathrm{kcal\,mol^{-1}}$, respectively, instead reflect scatter
among individual molecular errors. A species-resolved comparison of the B3LYP and r2SCAN atomization-energy errors is given in Appendix~\ref{app:species}. The deterioration after ML correction
therefore cannot be attributed simply to the removal of a systematic
offset.

In contrast to the atomization energies, the ML correction produces only
modest changes in the relative-energy benchmarks. For G2-IP, G2-EA,
ISO20, and ISO-C, the mean absolute changes in MAD across the four parent
functionals are only $0.39$, $0.34$, $0.09$, and
$0.47~\mathrm{kcal\,mol^{-1}}$, respectively. This is substantially
smaller than the corresponding changes for the atomization-energy sets:
$2.75$, $9.89$, and $20.08~\mathrm{kcal\,mol^{-1}}$ for G2-AE, Alk-AE,
and P6-AE, respectively. Thus, the dominant effect of the ML correction
is on absolute atomization energies, and this effect grows strongly with
the size of the molecular test set.

For the relative energies, the parent-functional differences are also
property dependent. SCAN and \rtscan{} give larger MADs than B3LYP for
both G2-IP and G2-EA: $6.47$ and $6.18~\mathrm{kcal\,mol^{-1}}$ versus
$4.92~\mathrm{kcal\,mol^{-1}}$ for ionization potentials, and $5.88$ and
$5.36~\mathrm{kcal\,mol^{-1}}$ versus $4.33~\mathrm{kcal\,mol^{-1}}$ for
electron affinities. These differences are consistent with the greater
sensitivity of these properties to one-electron errors and
self-interaction effects. A species-resolved comparison of B3LYP and
\rtscan{} is given in Appendix~\ref{app:species}.

\begin{table*}
\centering
\caption{CCSD(T) reference atomization energies and signed atomization-energy
errors for the three training molecules. The CCSD(T) reference values are
given in $\mathrm{kcal\,mol^{-1}}$, while all other entries are signed
errors relative to CCSD(T), in $\mathrm{kcal\,mol^{-1}}$. The final row
gives the mean signed error over the three training molecules.}
\label{tab:training}
\begin{ruledtabular}
\begin{tabular}{lccccccccc}
Species
& CCSD(T)
& PBE
& ML-PBE
& B3LYP
& ML-B3LYP
& SCAN
& ML-SCAN
& \rtscan
& \mlrtscan \\
\hline
H$_2$O
& 233.0
& $+0.01$
& $-0.66$
& $-4.23$
& $+0.07$
& $-5.81$
& $-2.84$
& $-4.73$
& $-3.55$ \\

C$_2$H$_2$
& 405.3
& $+9.00$
& $+6.55$
& $-4.26$
& $-0.64$
& $-4.86$
& $-0.00$
& $-3.64$
& $-0.20$ \\

SO$_2$
& 260.9
& $+17.99$
& $+15.11$
& $-14.15$
& $-0.27$
& $-4.48$
& $-0.02$
& $-1.94$
& $+0.03$ \\
\hline
Mean
& ---
& $+9.00$
& $+7.00$
& $-7.55$
& $-0.28$
& $-5.05$
& $-0.95$
& $-3.44$
& $-1.24$
\end{tabular}
\end{ruledtabular}
\end{table*}

\subsection{Training}
\label{sec:training}

Table~\ref{tab:training} compares the signed atomization-energy errors of
the three training molecules before and after ML correction. The effect of
the optimization differs substantially among the parent functionals. For
PBE, the mean signed error decreases only from
$+9.00$ to $+7.00~\mathrm{kcal\,mol^{-1}}$, with the individual errors for
C$_2$H$_2$ and SO$_2$ remaining at $+6.55$ and
$+15.11~\mathrm{kcal\,mol^{-1}}$, respectively. In contrast, the
corresponding mean errors for B3LYP, SCAN, and \rtscan{} are reduced from
$-7.55$, $-5.05$, and $-3.44~\mathrm{kcal\,mol^{-1}}$ to
$-0.28$, $-0.95$, and $-1.24~\mathrm{kcal\,mol^{-1}}$, respectively.
For these three parent functionals, the errors of C$_2$H$_2$ and SO$_2$
are reduced to values close to zero, whereas the H$_2$O error remains
more persistent. For example, the H$_2$O error changes from
$-4.23$ to $+0.07~\mathrm{kcal\,mol^{-1}}$ for B3LYP, but remains
$-2.84$ and $-3.55~\mathrm{kcal\,mol^{-1}}$ for SCAN and \rtscan{},
respectively.

The training results therefore show that the ML optimization reduces the
overall training error, but does not fit the three molecules uniformly for
all parent functionals. In particular, the substantially smaller
improvement obtained for PBE indicates that the optimization is not simply
driving every training target to zero. Because each model is obtained from
a single optimization run with a finite optimization budget, the remaining
errors cannot be unambiguously attributed to either a compromise among the
training targets or incomplete convergence of the optimizer.

For \rtscan{}, the final loss is $5.550\times10^{-3}$, compared with
$1.51\times10^{-2}$ for the uncorrected parent, corresponding to a
reduction by a factor of approximately $2.7$. For B3LYP, the corresponding
reduction factor is approximately $22$. Thus, the optimization produces a
substantially larger relative reduction in the loss for B3LYP than for
\rtscan{}. This difference is consistent with the smaller initial
training-set errors of \rtscan{}, although the single optimization run
does not allow us to determine whether the resulting parameters represent
the global minimum of the loss function.

To assess whether the ML correction adversely affects the one-electron
limit, we additionally consider the hydrogen atom, which is not included
in the training set. For a one-electron system, the exact
exchange--correlation energy exactly cancels the Hartree self-interaction,
giving a total ground-state energy of $-0.5$~Ha for the hydrogen atom.
Plain \rtscan{} gives an error of
$+0.046~\mathrm{kcal\,mol^{-1}}$ relative to this value, compared with
$-1.352~\mathrm{kcal\,mol^{-1}}$ for B3LYP. After applying the ML
correction, the \rtscan{} error becomes
$-0.062~\mathrm{kcal\,mol^{-1}}$. Although the sign changes, the
magnitude remains small, indicating that the ML correction does not
substantially worsen the one-electron error for the hydrogen atom.

\begin{figure*}%[t]
\includegraphics[width=\textwidth]{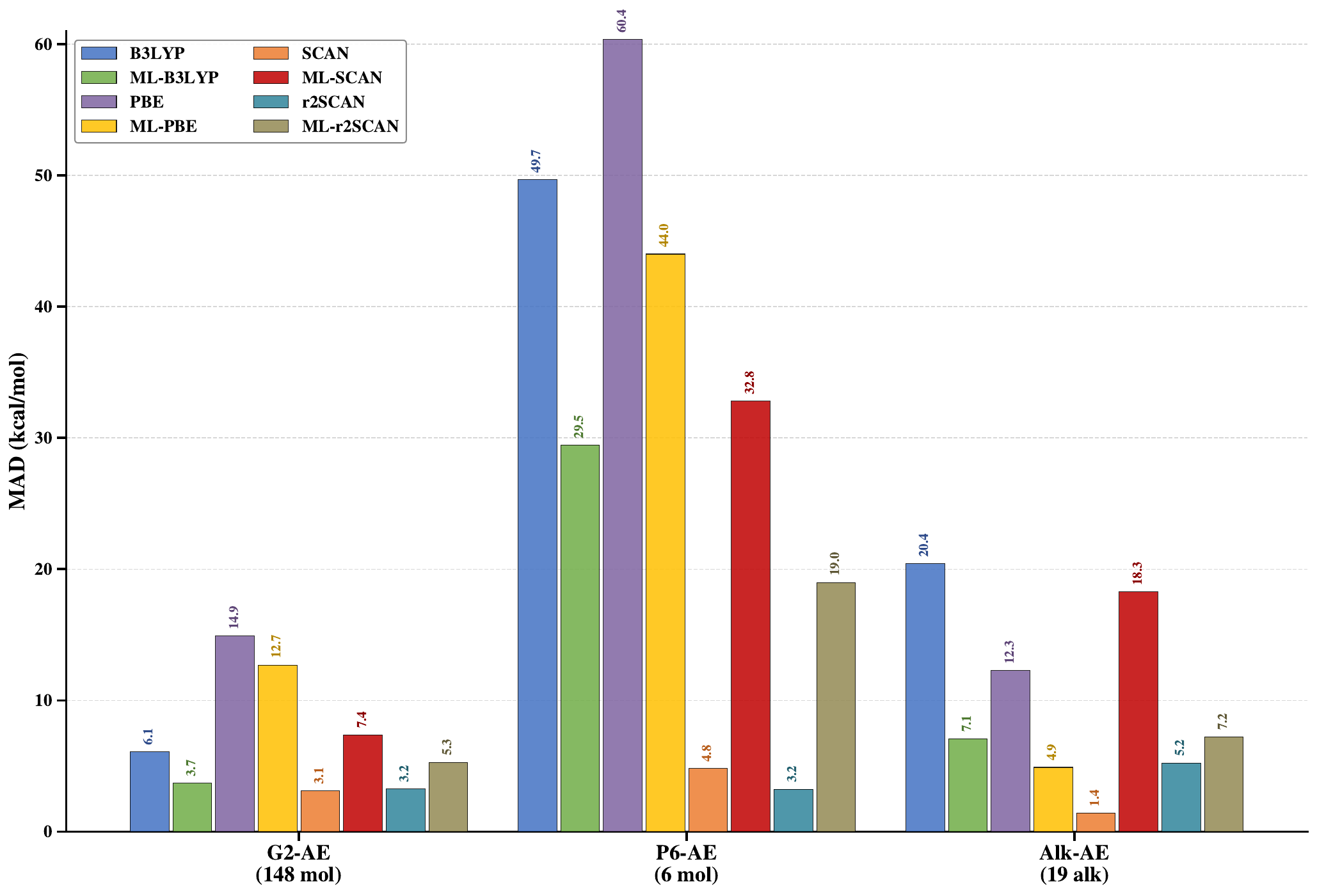}
\caption{Mean absolute deviation from CCSD(T) for absolute atomization
energies, Def2-TZVPD, for the four parent functionals and their
ML-corrected counterparts. The three sets are ordered left to right by
increasing typical molecular size. The correction improves B3LYP and PBE,
whose parent errors grow with system size, and degrades SCAN and \rtscan,
whose parent errors do not; the degradation grows with the size of the set.}
\label{fig:ae}
\end{figure*}

\begin{figure*}[t]
\includegraphics[width=\textwidth]{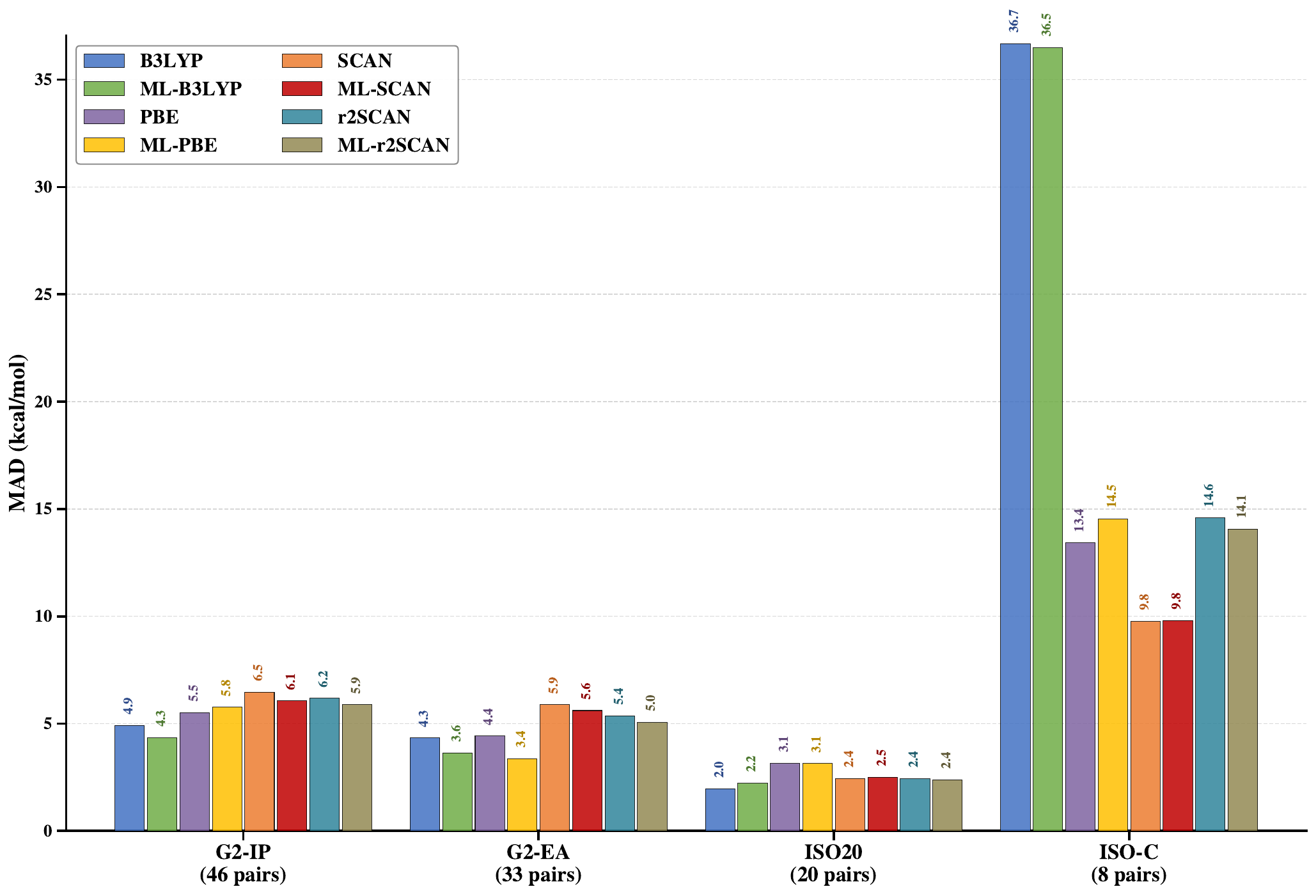}
\caption{Mean absolute deviation from CCSD(T) for relative energies, in
which the two states compared have the same composition. Unlike
Fig.~\ref{fig:ae}, the ML correction leaves every bar essentially unchanged
for every parent. The large ISO-C errors are pre-existing failures on
strained carbon cages, unaffected by the correction.}
\label{fig:rel}
\end{figure*}

\subsection{Size-dependent contribution of the learned correction}
\label{sec:extensive}

We next examine how the ML correction changes the atomization-energy error
with molecular size. We define
\begin{equation}
\Delta_i =
\varepsilon_i^{\mathrm{ML}}
-
\varepsilon_i^{\mathrm{parent}},
\end{equation}
where $\varepsilon_i$ is the signed atomization-energy error for species
$i$. Thus, $\Delta_i$ measures the contribution of the ML correction to the
error, independent of the error already present in the parent functional.

Along the $n$-alkane series, consisting of CH$_4$, C$_2$H$_6$, C$_3$H$_8$,
and the two C$_4$H$_{10}$ isomers, $\Delta_i$ varies approximately
linearly with molecular size. The corresponding slopes are listed in
Table~\ref{tab:slopes}, with $|r|\geq0.9999$ for all four parent
functionals. To examine whether this behavior extends beyond the alkane
series, we consider a broader set of 29 molecules containing hydrocarbons,
oxygen- and nitrogen-containing heterocycles, and small inorganic
molecules. For this set, we obtain
\begin{equation}
\Delta =
0.798\,N_{\mathrm{atoms}} - 0.267
~\mathrm{kcal\,mol^{-1}}
\qquad (r=0.935)
\end{equation}
for \mlrtscan{} and
\begin{equation}
\Delta =
1.133\,N_{\mathrm{atoms}} + 0.640
~\mathrm{kcal\,mol^{-1}}
\qquad (r=0.926)
\end{equation}
for ML-SCAN. The approximately linear dependence on the number of atoms
shows that the learned correction contains a systematic size-dependent
component. We use this terminology rather than claiming formal
size-extensivity of the ML functional, since the present analysis is based
on finite molecular test sets.

This size dependence is not explicitly imposed by the training objective in
Eq.~\eqref{eq:loss}. The training set contains only three molecules with
three or four atoms, and the loss function contains no explicit molecular
size variable. The observed scaling therefore emerges from the learned
semilocal correction and its application to larger systems.

\paragraph{Relation to the parent functional.}

The size-dependent contribution of the ML correction has opposite signs to
the corresponding size-dependent errors of PBE and B3LYP. For PBE, the
atomization-energy error along the alkane series has a slope of
$+3.469~\mathrm{kcal\,mol^{-1}}$ per CH$_2$ unit, whereas the ML correction
has a slope of $-1.444~\mathrm{kcal\,mol^{-1}}$ per CH$_2$ unit. The
correction therefore partially compensates for the size-dependent error of
PBE, reducing its slope by approximately $42\%$. For B3LYP, the
corresponding slopes are $-3.764$ and
$+2.384~\mathrm{kcal\,mol^{-1}}$ per CH$_2$ unit, giving a reduction of
approximately $63\%$ in the magnitude of the parent slope.

For SCAN and \rtscan{}, the ML contribution has the same positive sign in
both cases, with slopes of $+3.398$ and
$+2.184~\mathrm{kcal\,mol^{-1}}$ per CH$_2$ unit, respectively. The parent
slopes are only $+0.486$ and
$-0.316~\mathrm{kcal\,mol^{-1}}$ per CH$_2$ unit. Thus, unlike for PBE and
B3LYP, the ML correction does not compensate for a substantial
size-dependent error already present in the parent meta-GGAs. Instead, it
introduces a much larger size-dependent contribution.

The mean signed errors of the three training molecules provide another
comparison. Their values for PBE, B3LYP, SCAN, and \rtscan{} are
$+9.00$, $-7.55$, $-5.05$, and
$-3.44~\mathrm{kcal\,mol^{-1}}$, respectively, whereas the corresponding
ML slopes along the alkane series are $-1.444$, $+2.384$, $+3.398$, and
$+2.184~\mathrm{kcal\,mol^{-1}}$ per CH$_2$ unit. The signs are opposite
for all four parent functionals. However, with only four parent
functionals, this observation is insufficient to establish a general
relation between the training-set bias and the size-dependent behavior.
Consistently, the magnitude of the ML slope shows essentially no
correlation with either the magnitude of the training-set error
($r=0.03$) or the reduction in the training loss ($r=0.01$). Thus, the
present data do not establish whether the size-dependent contribution is
primarily determined by compensation of parent-functional errors or by the
semilocal form learned by the NN.

\paragraph{Connection to benchmark accuracy.}

The different size dependence of the parent functionals is reflected in
their G2-AE performance after ML correction. For the four parent
functionals, the change in G2-AE MAD can be fitted to the magnitude of the
parent slope as
\begin{equation}
\Delta\mathrm{MAD}
\approx
-1.69\,\left|\mathrm{slope}_{\mathrm{parent}}\right|
+3.89~\mathrm{kcal\,mol^{-1}},
\label{eq:breakeven}
\end{equation}
with the fitted line crossing zero at approximately
$|\mathrm{slope}_{\mathrm{parent}}|=2.3~\mathrm{kcal\,mol^{-1}}$ per
CH$_2$ unit. The ML correction reduces the G2-AE MAD for PBE from
$14.91$ to $12.81~\mathrm{kcal\,mol^{-1}}$ and for B3LYP from
$6.09$ to $3.69~\mathrm{kcal\,mol^{-1}}$. In contrast, the MAD increases
from $3.11$ to $7.51~\mathrm{kcal\,mol^{-1}}$ for SCAN and from
$3.25$ to $5.36~\mathrm{kcal\,mol^{-1}}$ for \rtscan{}.

Because the fit in Eq.~\eqref{eq:breakeven} contains only four data points,
it should not be interpreted as a predictive relation or as evidence for a
universal threshold. Rather, it summarizes the trend observed for the four
parent functionals: the ML correction improves the atomization-energy
accuracy of the two parents with substantial size-dependent errors, while
it degrades the two meta-GGAs whose corresponding parent errors are much
smaller.

\paragraph{Residual size dependence.}

The ML correction does not completely remove the size-dependent error for
any of the four parent functionals. For PBE and B3LYP, the residual slopes
of the ML-corrected functionals are $+2.025$ and
$-1.380~\mathrm{kcal\,mol^{-1}}$ per CH$_2$ unit, respectively. Thus, the
correction reduces, but does not eliminate, the size dependence already
present in these parent functionals. For SCAN and \rtscan{}, the residual
slopes increase to $+3.884$ and
$+1.868~\mathrm{kcal\,mol^{-1}}$ per CH$_2$ unit, respectively, compared
with only $+0.486$ and $-0.316~\mathrm{kcal\,mol^{-1}}$ for the parent
functionals. The ML correction therefore introduces a substantially larger
size-dependent error in the two meta-GGAs.

\paragraph{Transfer to larger molecules.}

We further test whether the size-dependent behavior identified above
persists in larger molecular benchmarks. The Alk-AE set contains 19 alkanes,
while P6-AE contains six larger molecules. The results are summarized in
Fig.~\ref{fig:ae}. For SCAN, the same deterioration observed in the
molecular-series analysis becomes substantially larger for these test sets:
the MAD increases from $1.42$ to $18.27~\mathrm{kcal\,mol^{-1}}$ for Alk-AE
and from $4.81$ to $32.80~\mathrm{kcal\,mol^{-1}}$ for P6-AE after applying
the ML correction. For \rtscan{}, the corresponding changes are from
$5.24$ to $7.24~\mathrm{kcal\,mol^{-1}}$ for Alk-AE and from $3.24$ to
$18.99~\mathrm{kcal\,mol^{-1}}$ for P6-AE. In contrast, the ML correction
improves both PBE and B3LYP on the larger test sets, reducing the Alk-AE MADs from 12.27 to 4.90~kcal\,mol$^{-1}$ for PBE and from 20.42 to 7.09~kcal\,mol$^{-1}$ for B3LYP.

The mean absolute change produced by the ML correction across the four
parent functionals increases from $2.75~\mathrm{kcal\,mol^{-1}}$ for G2-AE
to $9.89~\mathrm{kcal\,mol^{-1}}$ for Alk-AE and
$20.08~\mathrm{kcal\,mol^{-1}}$ for P6-AE. The increasing magnitude of the
ML-induced change with molecular size is consistent with the systematic
size-dependent contribution identified from the molecular-series analysis.

\begin{table*}%[t]
\centering
\caption{Linear size dependence of the atomization-energy errors along the
$n$-alkane series (CH$_4$, C$_2$H$_6$, C$_3$H$_8$, and the two
C$_4$H$_{10}$ isomers). All slopes are in
$\mathrm{kcal\,mol^{-1}}$ per CH$_2$ unit. The column
$\Delta$ gives the slope of the change in signed error produced by the ML
correction, while the residual slope is that of the corresponding
ML-corrected functional. The column ``Compensated'' gives
$-\mathrm{slope}(\Delta)/\mathrm{slope(parent)}$ and is reported only when
the magnitude of the parent slope is sufficiently large for the ratio to
be meaningful. The final column gives the Pearson correlation coefficient
for the linear fit of $\Delta$ versus molecular size.}
\label{tab:slopes}
\begin{ruledtabular}
\begin{tabular}{lccccc}
Functional
& Parent slope
& $\Delta$ slope
& Residual slope
& Compensated
& $r(\Delta)$ \\
\hline
PBE
& $+3.469$
& $\bm{-1.444}$
& $+2.025$
& $42\%$
& $-0.99986$ \\

B3LYP
& $-3.764$
& $\bm{+2.384}$
& $-1.380$
& $63\%$
& $+1.00000$ \\

SCAN
& $+0.486$
& $+3.398$
& $+3.884$
& --- 
& $+1.00000$ \\

\rtscan
& $-0.316$
& $+2.184$
& $+1.868$
& ---
& $+1.00000$ \\
\end{tabular}
\end{ruledtabular}
\end{table*}

\subsection{Where the extensive term resides}
\label{sec:where}

The analysis so far establishes that the correction is extensive without
saying what part of the density carries it. We therefore evaluated each
trained network on the converged parent densities of CH$_4$,
C$_2$H$_6$, C$_3$H$_8$ and the constituent atoms, and decomposed the
resulting contribution to the atomisation energy by region of space and by
descriptor value. No ML self-consistent calculation is involved.

\paragraph{The effect is purely an energy-density effect.}
The non-self-consistent estimate reproduces the self-consistent per-CH$_2$
slope almost exactly for every parent: $+2.381$ against $+2.384$ (B3LYP,
100\,\%), $+2.136$ against $+2.184$ (\rtscan, 98\,\%), $+3.402$ against
$+3.398$ (SCAN, 100\,\%) and $-1.438$ against $-1.444$ (PBE, 100\,\%).
Relaxation of the density under the ML potential contributes nothing
measurable to the extensivity; the term arises entirely from evaluating
$\Delta\epsilon$ on a density the parent would have produced anyway. This has two consequences. In practice, candidate modifications can be screened on stored densities without any SCF. Conceptually, a remedy must act on the energy expression or on the loss, not on the potential.

\paragraph{It lives in the bonding region.}
Table~\ref{tab:where} decomposes the CH$_4 \rightarrow$ C$_2$H$_6$ increment
by electron density. For B3LYP, SCAN and \rtscan{} the bonding bin
$0.1 < \rho < 1$ carries essentially the whole effect. In descriptor terms
the increment sits at $s \approx 0.1$--$0.6$, and spatially at $0.5$--$2$
bohr from the nearest nucleus: the C--H and C--C bond density. The
correction behaves as a fixed energy per bond, and a CH$_2$ unit adds a
fixed number of bonds.

The slowly varying region contributes almost nothing: $s < 0.1$ accounts for
1.9\,\%, 2.1\,\% and 3.0\,\% of the increment for the three parents. This
excludes an otherwise natural class of remedy, discussed in
Sec.~\ref{sec:conclusions}.

\paragraph{PBE is a different mechanism.}
Its increment is dominated not by the bonding region but by the carbon core,
the bin $10 < \rho < 100$ carrying 224\,\% of the total and the bonding
region partially opposing it. The core contribution scales as
$2.011\times$ from CH$_4$ to C$_2$H$_6$, that is, exactly per carbon atom. The magnitude of the correction on the isolated carbon atom is
correspondingly extreme: $-149.7$~\kcal{} for PBE against $-0.9$, $-2.9$ and
$-6.9$~\kcal{} for \rtscan, B3LYP and SCAN.

This is the pathology that the total-energy term of Eq.~\eqref{eq:loss} was
introduced to suppress: an atomisation energy reproduced through
near-cancellation of two very large total-energy errors. The two mechanisms are therefore distinct: a per-bond valence term for B3LYP, SCAN and \rtscan, and a per-core-atom term for PBE. Both are extensive, and both are invisible to a loss function built from three small molecules.

\begin{table*}%[t]
\centering
\caption{Decomposition of the CH$_4 \rightarrow$ C$_2$H$_6$ increment in
the ML contribution to the atomization energy according to electron-density
range. Each entry gives the signed contribution from the indicated density
range as a percentage of the total ML contribution to the increment.
The decomposition is evaluated using the converged parent-functional
densities. The remaining contribution in each column arises from the
low-density region, $\rho<0.1$. Values exceeding $100\%$ or becoming
negative reflect cancellation between contributions from different
density ranges.}
\label{tab:where}
\begin{ruledtabular}
\begin{tabular}{lcccc}
$\rho$ range
& B3LYP
& SCAN
& \rtscan
& PBE \\
\hline
$0.1 < \rho < 1$
& $96\,\%$
& $109\,\%$
& $105\,\%$
& $-105\,\%$ \\

$1 < \rho < 10$
& $9\,\%$
& $0\,\%$
& $0\,\%$
& $-36\,\%$ \\

$10 < \rho < 100$
& $0\,\%$
& $-1\,\%$
& $0\,\%$
& $+224\,\%$ \\

$\rho > 100$
& $0\,\%$
& $0\,\%$
& $0\,\%$
& $+4\,\%$ \\
\end{tabular}
\end{ruledtabular}
\end{table*}

\begin{figure*}%[t]
\includegraphics[width=\textwidth]{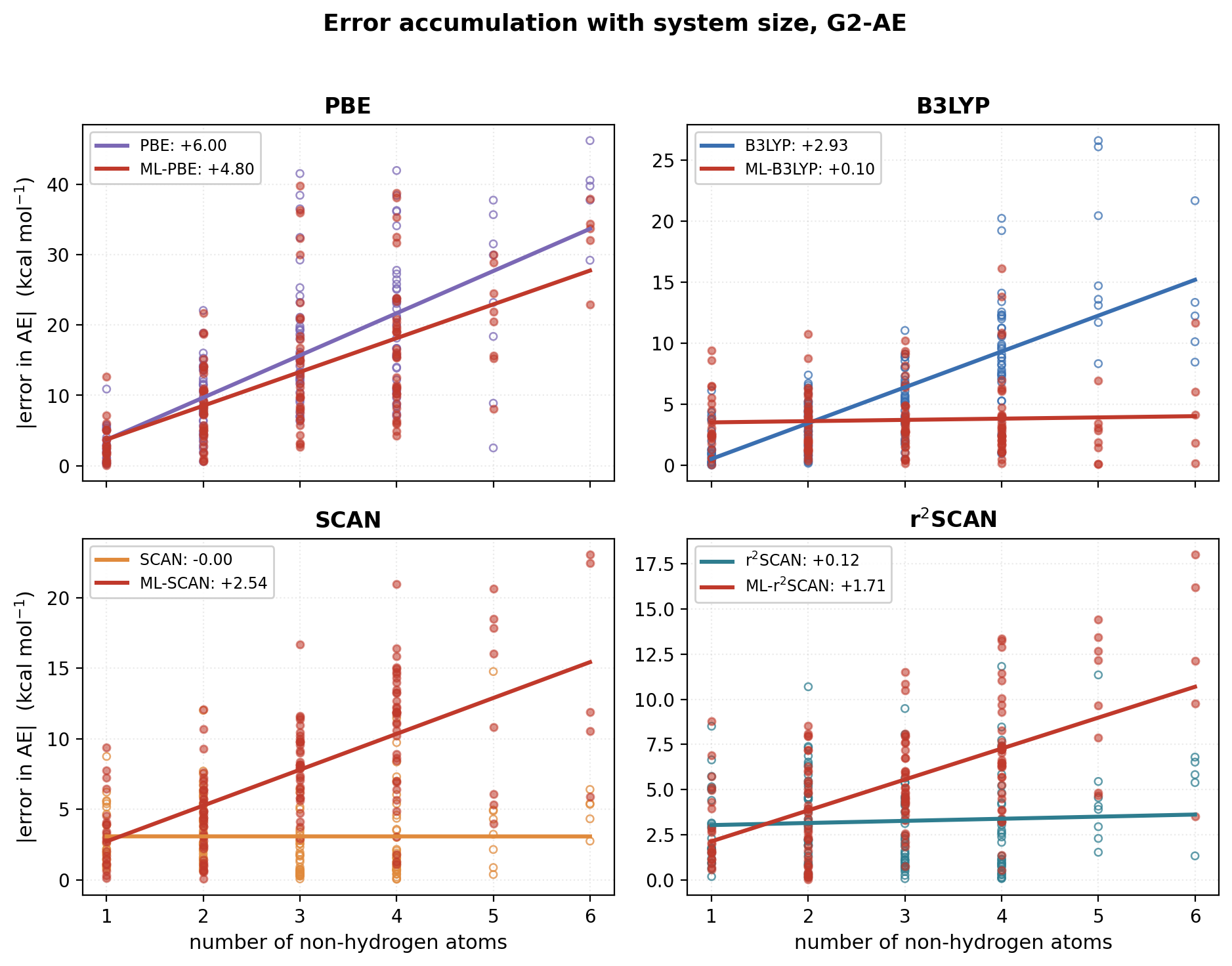}
\caption{Absolute atomization-energy error relative to CCSD(T) as a function
of the number of non-hydrogen atoms for the four parent functionals (open
circles) and ML-corrected counterparts (filled circles). From left to
right and top to bottom, the panels show PBE, B3LYP, SCAN, and \rtscan{}.
Solid lines are linear fits to the absolute errors, with the corresponding
slopes given in the legends in $\mathrm{kcal\,mol^{-1}}$ per non-hydrogen
atom. The G2-AE data contain 148 molecules for the parent functionals and
145 molecules for the ML-corrected functionals because the three training
molecules are excluded from the latter. The B3LYP panel provides a
comparison with Fig.~3 of Ref.~\cite{wang2023}.
}
\label{fig:accum}
\end{figure*}

\paragraph{Comparison with the error-accumulation analysis of
Ref.~\cite{wang2023}.} Figure~\ref{fig:accum} repeats the error-accumulation
analysis of Ref.~\cite{wang2023} for all four parents. For B3LYP the slope
falls from 2.93 to 0.10~\kcal{} per non-hydrogen atom, against the published
2.1 and 0.5; their fit also included the G3-HOF set, which may account for part of the difference in both the baseline and the residual. PBE shows the same
direction, from 6.00 to 4.80. For the two meta-GGAs the behaviour reverses:
plain SCAN and \rtscan{} show almost no growth of error with size (slopes
0.00 and 0.12), and the correction introduces it, 2.54 for ML-SCAN and 1.71
for \mlrtscan. The reduction in error accumulation reported for ML-B3LYP is
therefore not a property of the correction alone. It appears only when the
parent already has an error that grows with size, and where the parent has
none, the correction creates one. On the same axis, the earlier pointwise
model of Ref.~\cite{wang2022} suppressed extensivity by roughly 69\,\%,
against 63\,\% for B3LYP and 42\,\% for PBE here. The two quantities differ (theirs is the XC-energy error per non-hydrogen
atom on G3-HOF, ours the atomisation-energy error per CH$_2$), so the
comparison indicates direction only. Because this figure plots
absolute errors, it cannot show the sign of the learned term; that the
correction is negative for PBE and positive for B3LYP is established by the
signed fits of Table~\ref{tab:slopes}.

\subsection{Confirmation from the ion and isomerization benchmarks}
\label{sec:ions}

The small effect of the ML correction on ionization potentials and electron
affinities provides an important contrast to its much larger effect on
absolute atomization energies. Similar behavior was reported for the
pointwise ML correction of Ref.~\cite{wang2022}, where the weak response of
the ionic properties was initially discussed in connection with the
neutral-only training set. The present results show that the small changes
in the ionization and electron-affinity benchmarks persist across all four
parent functionals and therefore do not depend simply on the presence or
absence of ionic species in the training set.

Table~\ref{tab:cancel} quantifies the change in the signed error produced
by the ML correction. For the meta-GGA parents, the mean absolute changes
in the G2-IP and G2-EA errors are only $0.87$ and $0.37~\mathrm{kcal\,mol^{-1}}$
for SCAN and $0.47$ and $0.32~\mathrm{kcal\,mol^{-1}}$ for \rtscan{},
respectively. The largest individual changes are $2.10$ and
$1.61~\mathrm{kcal\,mol^{-1}}$ for SCAN and $1.22$ and
$0.99~\mathrm{kcal\,mol^{-1}}$ for \rtscan{}. The per-species error
patterns are also highly correlated before and after ML correction, with
Pearson coefficients of $0.9980$ and $0.9976$ for SCAN and
$0.9994$ and $0.9991$ for \rtscan{} for the G2-IP and G2-EA sets,
respectively. Thus, the ML correction has only a minor effect on the
relative ordering of the ionization-potential and electron-affinity errors
of the meta-GGAs.

The contrast with atomization energies is substantially larger. For the
G2-AE set, the mean absolute changes in the signed error are
$3.60$, $8.73$, $5.99$, and $3.85~\mathrm{kcal\,mol^{-1}}$ for PBE, B3LYP,
SCAN, and \rtscan{}, respectively, compared with
$0.65$, $2.02$, $0.87$, and $0.47~\mathrm{kcal\,mol^{-1}}$ for G2-IP.
The corresponding ratios are approximately $6$, $4$, $7$, and $8$.
Relative to G2-EA, the ratios are approximately $5$, $10$, $16$, and
$12$. The much larger response of atomization energies therefore cannot be
explained simply by the ML correction being generally small. Rather, its
effect depends strongly on the type of energy difference being considered.

There is also little evidence for a systematic size dependence within the
ionization-potential benchmarks. For \rtscan{}, for example, the change in
the CH$_2$ ionization-potential error is $+0.66~\mathrm{kcal\,mol^{-1}}$,
whereas the corresponding change for C$_6$H$_5$CH$_3$ is
$+0.53~\mathrm{kcal\,mol^{-1}}$, despite the substantial difference in
molecular size. A particularly direct comparison is provided by benzene,
for which both atomization and ionization energies are available. The
atomization-energy error changes by $-8.47$, $+16.65$, and
$+11.20~\mathrm{kcal\,mol^{-1}}$ for PBE, SCAN, and \rtscan{}, respectively,
whereas the corresponding ionization-potential changes are only
$-0.34$, $+0.93$, and $+0.58~\mathrm{kcal\,mol^{-1}}$. The magnitude of the
atomization-energy response is therefore larger by factors of approximately
25, 18, and 19, respectively.

Ionization potentials and electron affinities, however, change the electron
number between the two states. Isomerization energies provide a
complementary test in which both states have the same composition, charge,
and number of electrons, but differ in molecular structure. The ML
correction again has only a small effect. For the 20 ISO20 isomerization
energies, the changes in MAD are $-0.01$, $+0.26$, $+0.06$, and
$-0.04~\mathrm{kcal\,mol^{-1}}$ for PBE, B3LYP, SCAN, and \rtscan{},
respectively. For the eight ISO-C isomerization energies, the corresponding
changes are $+1.11$, $-0.19$, $+0.05$, and
$-0.52~\mathrm{kcal\,mol^{-1}}$. Across these 16 relative-energy
comparisons, the mean absolute change is only
$0.32~\mathrm{kcal\,mol^{-1}}$, with a maximum of
$1.11~\mathrm{kcal\,mol^{-1}}$. Figure~\ref{fig:rel} shows this weak
response alongside the ionization-potential and electron-affinity results,
in contrast to the strong size-dependent changes in the atomization-energy
benchmarks shown in Fig.~\ref{fig:accum}.

Taken together, these results indicate that the ML correction contributes
much more strongly when comparing a molecule with its separated atoms than
when comparing two closely related states of the same molecular system.
This behavior is consistent with a local or semilocal correction whose
contribution can accumulate in absolute energies while partially cancelling
between chemically related states. The ionization and isomerization
benchmarks are not fully independent tests, since both involve substantial
cancellation of local contributions. Nevertheless, their combined behavior
supports the interpretation that the dominant ML-induced error is associated
with the change in density distribution between the molecular and reference
states rather than simply with changes in electron number.

Two classes of pre-existing error remain largely unaffected by the ML
correction. First, the ISO-C errors are large for all four parent
functionals, ranging from $9.8$ to $36.7~\mathrm{kcal\,mol^{-1}}$, and
change little after ML correction. These errors therefore represent
limitations of the parent functionals for the strained carbon-cage
isomerizations rather than errors that are substantially corrected by the
present ML construction. Second, individual systems dominate the G2-IP and
G2-EA errors. In particular, CN has errors of $+34.42$ and
$+33.90~\mathrm{kcal\,mol^{-1}}$ before and after ML correction,
respectively, while the corresponding C$_2$ G2-EA errors are
$+22.49~\mathrm{kcal\,mol^{-1}}$ in both cases. These systems involve
strong multireference character, for which single-reference CCSD(T)
reference values may themselves be less reliable. Excluding them reduces
the G2-IP MAD from $5.55$ to $5.25~\mathrm{kcal\,mol^{-1}}$ and the G2-EA
MAD from $4.83$ to $4.50~\mathrm{kcal\,mol^{-1}}$ for the corresponding
plain and ML-corrected calculations . The negligible changes for CN and
C$_2$ further indicate that the present semilocal correction does not
address errors associated with static correlation, which requires a
different treatment~\cite{becke2013static}.

\begin{table}%[t]
\centering
\caption{Effect of the ML correction on the signed errors for the
different benchmark properties. The change in signed error is defined as
$\Delta=\varepsilon^{\mathrm{ML}}-\varepsilon^{\mathrm{parent}}$, where
$\varepsilon$ is the signed error relative to the CCSD(T) reference. The
columns $\langle\Delta\rangle$, $\langle|\Delta|\rangle$, and
$\max|\Delta|$ give the mean signed change, mean absolute change, and
maximum absolute change, respectively. All values are in
$\mathrm{kcal\,mol^{-1}}$. The PBE G2-EA set contains 32 rather than
33 systems because the C$_2^-$ anion does not converge for either PBE or
ML-PBE and is therefore excluded from the comparison.}
\label{tab:cancel}
\begin{ruledtabular}
\begin{tabular}{llcccc}
Parent & Property & $N$ &
$\langle\Delta\rangle$ &
$\langle|\Delta|\rangle$ &
$\max|\Delta|$ \\
\hline
\multirow{3}{*}{PBE}
& G2-AE & 145 & $-2.70$ & $3.60$ & $8.47$ (C$_6$H$_6$) \\
& G2-IP & 46  & $-0.42$ & $0.65$ & $2.05$ (He) \\
& G2-EA & 32  & $-0.78$ & $0.79$ & $1.62$ (NCO) \\[2pt]

\multirow{3}{*}{B3LYP}
& G2-AE & 145 & $+8.73$ & $8.73$ & $22.03$ (SiF$_4$) \\
& G2-IP & 46  & $+1.88$ & $2.02$ & $7.88$ (Ar) \\
& G2-EA & 33  & $+0.24$ & $0.85$ & $3.31$ (CH$_3$S) \\[2pt]

\multirow{3}{*}{SCAN}
& G2-AE & 145 & $+5.99$ & $5.99$ & $16.65$ (C$_6$H$_6$) \\
& G2-IP & 46  & $+0.85$ & $0.87$ & $2.10$ (CO$_2$) \\
& G2-EA & 33  & $+0.26$ & $0.37$ & $1.61$ (NCO) \\[2pt]

\multirow{3}{*}{\rtscan}
& G2-AE & 145 & $+3.85$ & $3.85$ & $11.20$ (C$_6$H$_6$) \\
& G2-IP & 46  & $+0.42$ & $0.47$ & $1.22$ (Ar) \\
& G2-EA & 33  & $+0.27$ & $0.32$ & $0.99$ (NCO) \\
\end{tabular}
\end{ruledtabular}
\end{table}

\subsection{Consequences for the interpretation of ML-B3LYP}
\label{sec:reinterp}

Section~\ref{sec:extensive} allows the size dependence of the parent
functional and of the learned correction to be examined separately. In a
companion reproduction, the retrained ML-B3LYP model showed a decreasing
atomization-energy error with increasing molecular size, reaching
$1.71~\mathrm{kcal\,mol^{-1}}$ for large hydrocarbons compared with
$2.40~\mathrm{kcal\,mol^{-1}}$ for the published functional. This behavior
can be understood from the separate size-dependent contributions identified
here. B3LYP has a parent-functional slope of
$-3.764~\mathrm{kcal\,mol^{-1}}$ per CH$_2$ unit, while the learned
correction contributes a slope of
$+2.384~\mathrm{kcal\,mol^{-1}}$ per CH$_2$ unit. The two contributions
therefore partially compensate, leaving a residual slope of
$-1.380~\mathrm{kcal\,mol^{-1}}$ per CH$_2$ unit. The apparent improvement
with molecular size is thus consistent with partial cancellation between
two size-dependent contributions of opposite sign.

This observation changes how the performance of ML-B3LYP on larger
molecules should be interpreted. Its improved size behavior does not, by
itself, demonstrate that the neural network has learned a transferable
component of the exchange--correlation energy. Instead, the improvement is
consistent with compensation between the size-dependent error of B3LYP and
that introduced by the ML correction. The residual slope shows that this
compensation is incomplete, with approximately $63\%$ of the magnitude of
the B3LYP slope offset by the learned correction.

The comparison with SCAN and \rtscan{} further illustrates why the parent
functional matters. Their parent slopes are only
$+0.486$ and $-0.316~\mathrm{kcal\,mol^{-1}}$ per CH$_2$ unit, whereas the
corresponding ML contributions are $+3.398$ and
$+2.184~\mathrm{kcal\,mol^{-1}}$ per CH$_2$ unit. The ML correction
therefore introduces a much larger size-dependent contribution than is
present in either parent. For PBE, in contrast, the ML contribution has
the opposite sign to the parent slope and partially compensates for it.
These results indicate that the performance of the ML correction depends
strongly on the size-dependent error already present in the parent DFA.
They do not, however, establish a universal criterion for predicting the
performance of the correction from the parent slope alone.

More generally, the present results suggest that the success of an ML-DFA
correction should not be assessed solely from its performance on a single
parent functional or from its behavior on larger molecules. A correction
can improve the apparent size dependence of one parent through error
compensation while introducing a new size-dependent contribution when
applied to another parent. This distinction is particularly important for
SCAN and \rtscan{}, for which the parent functionals already exhibit weak
size dependence in the present benchmark.

\subsection{Reliability of the ML-PBE model}
\label{sec:notes}

All four parent functionals were treated using the same training and
evaluation protocol, allowing their different responses to the ML
correction to be compared directly. A specific caveat applies to ML-PBE.
Among the four retrained models, ML-PBE shows the largest discrepancy from
the published ML-PBE results: its per-species atomization-energy errors
differ from the published ML-PBE column by
$15.2~\mathrm{kcal\,mol^{-1}}$ in the G2-AE benchmark. Its absolute
atomization-energy performance should therefore be interpreted with
caution. This discrepancy is also consistent with the strong
density-region dependence of the ML-PBE correction discussed in
Sec.~\ref{sec:where} and with the behavior reported previously for ML-PBE
~\cite{wang2023}.

The size-dependent analysis is less sensitive to this discrepancy because
the slope is determined internally from the ML-PBE results obtained with
the present implementation. The ML-PBE correction has a slope of
$-1.444~\mathrm{kcal\,mol^{-1}}$ per CH$_2$ unit, opposite in sign to the
$+3.469~\mathrm{kcal\,mol^{-1}}$ per CH$_2$ unit slope of PBE. Thus, the
qualitative conclusion that the ML correction partially compensates for the
size-dependent PBE error does not depend on agreement with the published
ML-PBE atomization-energy values. Comparative notes on SCAN and r2SCAN, and a summary of the SCF convergence failures encountered in this work, are given in Appendix ~\ref{App:points}.

\section{Conclusions and Outlook}
\label{sec:conclusions}
% =====================================================================

We have examined the transferability of the semilocal ML-DFA correction of
Ref.~\cite{wang2023} by applying the same ML construction and training
protocol to four parent functionals, PBE, B3LYP, SCAN, and \rtscan{}. The
comparison reveals a systematic dependence of the ML correction on the
size-dependent error of the parent functional.

First, the ML correction develops a nearly linear size-dependent
contribution to the atomization-energy error along the $n$-alkane series,
with $|r|\geq0.9999$ for all four parent functionals. The broader
29-molecule analysis gives similarly strong correlations for ML-SCAN and
\mlrtscan{}. We therefore find a systematic size-dependent contribution
of the learned correction, although the present finite-molecule analysis
does not by itself establish formal size-extensivity of the ML functional.
For PBE and B3LYP, the learned contribution has the opposite sign to the
dominant size-dependent error of the parent and therefore partially
compensates for it. For SCAN and \rtscan{}, whose parent size dependence is
much smaller, the ML correction instead introduces a substantially larger
size-dependent contribution. This difference is reflected directly in the
G2-AE performance: the ML correction improves PBE and B3LYP but degrades
SCAN and \rtscan{}, with the effect becoming increasingly pronounced for
larger molecular test sets.

Second, the size-dependent behavior is not imposed explicitly by the training objective. The training set contains only three small molecules, and the loss function contains no molecular-size variable. Moreover, evaluating the trained correction on frozen parent densities reproduces approximately $98$--$100\%$ of the size-dependent slope obtained from the self-consistent calculations. Thus, the dominant contribution to the observed size dependence originates from the learned energy correction itself rather than from the subsequent SCF relaxation. The density decomposition further indicates that the relevant contribution is associated primarily with the bonding region of the molecular density. For PBE, the density dependence differs qualitatively from that of the other parent functionals and is associated with a strong core-region contribution,whereas for B3LYP, SCAN, and r$^2$SCAN it is dominated by the bonding region, $0.1 \leq \rho \leq 1$.

Third, the strong effect on atomization energies does not extend to the
relative-energy benchmarks considered here. Across the ionization
potential, electron-affinity, and isomerization benchmarks, the mean
absolute change produced by the ML correction is only
$0.32~\mathrm{kcal\,mol^{-1}}$, compared with
$10.91~\mathrm{kcal\,mol^{-1}}$ for the corresponding atomization-energy
entries. The same contrast is observed as the molecular size increases:
the mean absolute change in G2-AE, Alk-AE, and P6-AE is
$2.75$, $9.89$, and $20.08~\mathrm{kcal\,mol^{-1}}$, respectively.
These results indicate that the dominant ML-induced error is associated
with comparisons between densities of substantially different character,
as in molecular atomization, rather than with changes in electron number
alone. The weak response of the isomerization benchmarks provides an
additional control because the compared states have the same composition
and electron number.

The results also provide a different interpretation of the previously
reported size behavior of ML-B3LYP. Its improved performance for larger
molecules is consistent with partial cancellation between the
size-dependent error already present in B3LYP and the opposite
size-dependent contribution introduced by the ML correction. The
improvement therefore cannot, by itself, be taken as evidence that the
network has learned a transferable component of the exchange--correlation
functional. More generally, the present comparison demonstrates that the
performance of an ML correction cannot be separated from the error
structure of its parent DFA.

Several directions follow from these observations. One possibility is to
modify the training objective so that size-dependent errors are explicitly
constrained, for example by including at least one larger molecule in the
training set or by normalizing energetic errors with respect to molecular
size. Such modifications would test whether the observed behavior is
primarily controlled by the training objective. A related strategy is the
species weighting used in the pointwise construction of Ref.~\cite{an2025mitigating}.
Another possibility is to change the functional representation itself,
rather than the loss function, so that undesirable accumulation of local
errors is constrained during training. These alternatives can be
distinguished systematically through controlled retraining and ablation
studies.

The density decomposition also constrains possible functional
modifications. For B3LYP, SCAN, and \rtscan{}, only a small fraction of the
size-dependent contribution originates from the slowly varying
small-$s$ region: the $s<0.1$ region accounts for approximately $1.9\%$,
$2.1\%$, and $3.0\%$, respectively. The dominant contribution instead
arises at intermediate values of $s$, approximately
$0.1\lesssim s\lesssim0.6$. Consequently, modifications designed solely
to enforce the small-$s$ gradient expansion are unlikely to remove the
dominant source of the observed size dependence.

Finally, the present analysis suggests several direct tests of the
generality of these conclusions. Because most of the size-dependent
contribution is reproduced on frozen parent densities, the same
alkane-based diagnostic can be applied non-self-consistently to other
published ML-DFA constructions for which the model parameters are
available. In particular, comparisons with the pointwise construction of
Ref.~\cite{an2025mitigating} could help determine whether the choice of training
target is associated with improved size transferability. Retraining the
global-loss model with a different optimizer while keeping the loss
function and training data unchanged would provide a complementary test
of whether the observed behavior depends on the optimization procedure.
Together, these tests would help distinguish the roles of the training
objective, model representation, and optimization strategy in determining
the size transferability of ML corrections to density functionals.

% \section*{Dedication}

% Although none of us had the opportunity to meet Axel Becke personally, his
% scientific ideas have had a lasting influence on the way we think about
% density functional theory. The exchange functional of
% Ref.~\cite{becke88}, the adiabatic-connection mixing of
% Ref.~\cite{becke}, and the real-space analyses that followed from
% them~\cite{beckeroussel,beckeedgecombe} demonstrated how fundamental
% physical principles, mathematical insight, and careful construction can lead
% to methods that are both simple and remarkably effective. Two of the four parent functionals examined here descend directly from that work.

% We dedicate this work to the memory of Axel Becke, with sincere gratitude
% for his profound contributions to density functional theory. His work
% continues to influence the development of new ideas in electronic-structure
% theory and will remain an enduring part of the intellectual foundation of
% our field.

\begin{acknowledgments}
AB and KKM thank the computing facility at NISER for resources.
\end{acknowledgments}

\bibliography{references}
%\bibliographystyle{apsrev4-2}

% =====================================================================
% SUPPLEMENTAL MATERIAL
% ---------------------------------------------------------------------
% Everything below this line is material moved out of the main text.
% To split it into a separate document: cut from \clearpage to \end{document},
% paste into a new file with the same preamble, and delete this block.
% =====================================================================

\clearpage
\onecolumngrid
\appendix
\section*{Appendix}
\setcounter{table}{0}
\renewcommand{\thetable}{S\arabic{table}}
\setcounter{section}{0}

\section{Implementation and verification\label{App:implementaion}}

\paragraph{Particle-swarm coefficients.}
The coefficients $\omega=0.9$, $\phi_p=0.95$, $\phi_g=0.9$ are not exposed
as constructor arguments in \texttt{nevergrad} 0.4.3, which hard-codes the
SPSO2011 defaults; we recovered them by subclassing the runtime PSO class.
Seeding the swarm at the origin, where
$\Delta\epsilon^{\text{ML}}_{XC}\equiv 0$, reduces the model exactly to the
parent and so also verifies that every functional-injection call site has
been switched consistently.

\paragraph{Meta-GGA potential propagation.}
A passthrough test, in which the custom functional routine merely calls the
library implementation of \rtscan{} and returns it, reproduces the native
\rtscan{} energy of H$_2$O to $1.99\times10^{-13}$~Ha, confirming that the
$\tau$ derivative is applied as an operator on the orbitals rather than
dropped. The $\tau$ convention was fixed independently on the hydrogen atom,
a one-electron system for which
$\alpha=(\tau-\tau_W)/\tau^{\text{unif}}$ must vanish identically; the
computed $\alpha$ is zero to six decimal places at all $2.2\times10^4$ grid
points.

\section{Error structure of the plain functionals ~\label{App:AE-err} }

B3LYP's largest G2-AE failures are systematic and chemically localised: SiCl$_4$ $-26.6$, SiF$_4$ $-26.1$, CCl$_4$ $-20.4$, AlCl$_3$
$-20.2$~\kcal, all heavy-halide underbinding. \rtscan{} removes precisely
this class (SiCl$_4$ $+1.53$, CCl$_4$ $+3.90$, AlCl$_3$ $+0.86$) and
replaces it with smaller bidirectional scatter, the largest cases being
ClF$_3$ $+11.82$ and SiF$_4$ $-11.35$~\kcal.

\section{Per-species detail along the alkane series \label{app:species}}

\begin{table}[h]
\centering
\caption{Per-species detail underlying Table~\ref{tab:slopes} for the two
parents with large extensive errors, \kcal. Note the opposite signs of
$\Delta$ and the near-constant increments.}
\label{tab:alkdetail}
\begin{ruledtabular}
\begin{tabular}{lcccccc}
& \multicolumn{3}{c}{PBE} & \multicolumn{3}{c}{B3LYP} \\
\cline{2-4}\cline{5-7}
Species & parent & $\Delta$ & incr. & parent & $\Delta$ & incr. \\
\hline
CH$_4$        & $-0.60$  & $-1.02$ & ---     & $-0.64$  & $+3.03$  & ---    \\
C$_2$H$_6$    & $+3.51$  & $-2.51$ & $-1.49$ & $-3.48$  & $+5.41$  & $+2.38$\\
C$_3$H$_8$    & $+6.99$  & $-3.95$ & $-1.44$ & $-7.27$  & $+7.79$  & $+2.38$\\
C$_4$H$_{10}$ & $+10.30$ & $-5.40$ & $-1.45$ & $-11.24$ & $+10.18$ & $+2.39$\\
\end{tabular}
\end{ruledtabular}
\end{table}

\section{Notes on individual parents ~\label{App:points} }

\paragraph{SCAN versus \rtscan.}
The regularisation costs nothing here. Plain SCAN gives a G2-AE MAD of
3.11~\kcal{} against \rtscan's 3.25, with MSD $-0.10$ against $+0.15$; on
the ion sets \rtscan{} is marginally the better of the two. No evidence of
SCAN's documented numerical sensitivity~\cite{rscan} is visible at this grid
level for these molecules, and all 148 G2-AE species converged on plain DIIS
for both parents. SCAN is thus the better plain functional on G2-AE and the
worse ML-corrected one; this ordering is not explained by the parents'
errors on the training set, but tracks the magnitude of the learned
extensive term.

\paragraph{Convergence.}
Two ML-SCAN G2-IP pairs (CO$_2$ and CS$_2$) failed to converge within 100
SCF cycles. Together with the C$_2$ anion, which fails for both PBE and
ML-PBE and is excluded from the PBE G2-EA statistics, these are the only
convergence failures in the study. The ML-SCAN failures occur for the
ML-corrected functional only, where second-order SCF is unavailable because
the custom functional supplies no XC kernel (Sec.~\ref{sec:comp}).

\end{document}